\documentclass[prd,twocolumn,nofootinbib]{revtex4}

\usepackage{amsmath,amssymb,graphicx,subfigure}

\begin{document}

\preprint{\hepth{0606114} \\ UCB-PTH-06/09 \\ LBNL-60250}

\title{Eternal inflation: The inside story}

\author{Raphael Bousso, Ben Freivogel and I-Sheng Yang\footnote{bousso@lbl.gov, freivogel@berkeley.edu, 
jingking@berkeley.edu}}

\affiliation{Department of Physics and Center for Theoretical
Physics \\
University of California, Berkeley, CA 94720, U.S.A. \\
{\em and}\\
Lawrence Berkeley National Laboratory, Berkeley, CA 94720, U.S.A. }
\begin{abstract}%
  Motivated by the lessons of black hole complementarity, we develop a
  causal patch description of eternal inflation.  We argue that an
  observer cannot ascribe a semiclassical geometry to regions outside
  his horizon, because the large-scale metric is governed by the
  fluctuations of quantum fields. In order to identify what is within
  the horizon, it is necessary to understand the late time
  asymptotics. Any given worldline will eventually exit from eternal
  inflation into a terminal vacuum.  If the cosmological constant is
  negative, the universe crunches.  If it is zero, then we find that
  the observer's fate depends on the mechanism of eternal inflation.
  Worldlines emerging from an eternal inflation phase driven by
  thermal fluctuations end in a singularity.  By contrast, if eternal
  inflation ends by bubble nucleation, the observer can emerge into an
  asymptotic, locally flat region.  As evidence that bubble collisions
  preserve this property, we present an exact solution describing the
  collision of two bubbles.

\end{abstract}

\maketitle

\section{Introduction}

Theories with multiple metastable vacua tend to exhibit the
cosmological dynamics known as eternal
inflation~\cite{ColDel80,GutWei83,Vil83b,Lin86b}.  This is usually
described in terms of a spacetime far larger than the presently
visible region.  Every vacuum is realized, over and over, in different
regions of this ever-expanding universe.  A longstanding problem has
been to compute the probability that a given vacuum will be observed.

Empirically, the small nonzero value of the cosmological
constant~\cite{Rie98} suggests that the number of metastable vacua is
extremely large~\cite{Sak84,Wei87}.  On the theoretical side, string
theory appears to satisfy this requirement~\cite{BP,KKLT}.  Eternal
inflation populates the landscape.  Hence, the notorious problem of a
probability measure has received renewed interest.

\begin{figure*}
\begin{center}
\includegraphics[scale = .6]{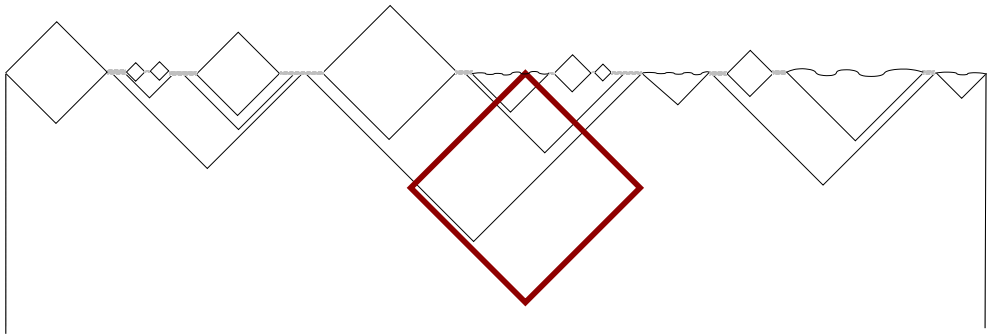}
\end{center}
\caption{A conformal diagram for the global geometry in FVEI.  The
  hats represent the future infinity of open FRW universes with
  vanishing cosmological constant.  Regions with negative cosmological
  constant end in a big crunch singularity (squiggly lines).  The part
  of the universe which remains in the inflationary phase has zero
  comoving volume but infinite physical volume.  The thick diamond
  shows an example of a causally connected region accessible to a
  single observer.  The local description developed here is confined
  to such regions.}
\label{fig-fveipenrose}
\end{figure*}

There are actually two kinds of eternal inflation.  False vacuum
eternal inflation (FVEI) is driven by the de~Sitter expansion of the
metastable vacua themselves.  They decay when bubbles of lower energy
vacua form spontaneously.  But inflation perdures because the bubbles
do not grow fast enough to catch up with the de Sitter expansion of
the false vacuum inside of which they formed~\cite{GutWei83}.  Thus
eventually all allowed transitions, no matter how suppressed, will
take place in different regions of the universe, and the landscape is
populated.  The global picture is that of an infinite universe
containing infinitely many bubbles, nested inside each other or
expanding in different places, occasionally colliding but never
percolating (Fig.~\ref{fig-fveipenrose}).  In some bubbles inflation
can end locally.  If the cosmological constant is negative, it ends in
a big crunch; if it vanishes, then an open FRW universe is produced.
If it is positive, it is still in a de~Sitter phase, and further
bubbles will eventually be produced.

Slow-roll eternal inflation (SREI) does not rely on local minima in
the potential.  Instead, it requires a potential $V$ in which a scalar
field $\phi$ classically rolls down slowly.  If $dV/d\phi\ll V^{3/2}$,
the quantum fluctuations of the scalar field dominate over its
classical evolution, and the field undergoes a kind of random
walk~\cite{Linde,GonLin87}.  In some regions of the inflating universe,
the field will wander into a regime where the classical evolution
dominates and inflation can end.  But there will always be other
regions in which the field fluctuates up and the inflationary
expansion continues.  If more than one final vacuum is accessible from
the SREI regime, or if certain moduli can be driven to different
values during SREI, one would again like to compute the probabilities
for different outcomes.

Intuitively, one might expect the probability to be proportional to
the fraction of the volume of the universe taken up by the vacuum in
question.  But this prescription is ill-defined.  In eternal inflation
there is no preferred foliation of spacetime into spatial slices, and
different choices lead to radically different answers~\cite{LinLin96}.
This is easily seen as follows.  In a region occupied by one vacuum,
there can be preferred slices, for example the comoving slices that
are orthogonal to the four-velocity of the cosmic fluid.  In our
region, those are the slices on which the universe appears homogeneous
and isotropic.  During ordinary slow-roll inflation, de~Sitter
invariance is broken by the time-dependence of the inflaton field, and
the hypersurfaces of constant inflaton field provide a preferred
slicing.

But there is no natural, unique way of extending such slices to cover
different vacua separated by an inflating region.  The problem is that
by assumption, any global slice through the geometry must include
regions that are in the eternally inflating phase.  In regions trapped
in a false vacuum, the geometry is locally de~Sitter invariant.  The
case of slow-roll eternal inflation is similar.  When quantum
fluctuations dominate over the classical slow-roll, hypersurfaces of
constant density will not be spacelike, restoring for all practical
purposes the local de~Sitter invariance.

But there are reasons to believe that a global description must be
inconsistent~\cite{SusTho93}.  In the context of black holes, it leads
to sharp paradoxes that are resolved only if we describe one observer
at a time.  This suggests more generally that we cannot talk
simultaneously about two observers who are forever out of causal
contact.  (Since they can never compare their experiences, this is not
a restriction on our ability to describe the world.)  Perhaps, then,
the ambiguities and paradoxes of eternal inflation are due, at least
in part, to our insistence on a global point of view.

The goal of this paper is to begin exploring a local description.
What does eternal inflation look like to a single observer?\footnote{A
  probability measure based on the local viewpoint is given in
  Ref.~\cite{Bou06}.}

We begin, in Sec.~\ref{sec-gedanken}, by asking whether a single
observer can predict the geometry of regions behind his event horizon.
This is not obviously impossible.  For example, we can predict with
some confidence the interior geometry of a black hole formed from
stellar collapse, even if we remain forever outside.  However, this
ability depends on the assumption that the gravitational backreaction
of quantum fluctuations is small.

To illustrate this point, we consider a thought experiment.  An
apparatus measures the spin of an electron {\em after\/} both electron
and apparatus fall into a black hole.  Thanks to standard decoherence
effects, the apparatus will show a definite outcome after the
measurement.  But the outside observer is ignorant of the outcome, and
thus cannot predict which of two macroscopically distinguishable
events will take place inside the black hole.  But if the pointer of
the apparatus is very massive, the metric inside the black hole will
depend on its position.  Hence, the outside observer cannot predict a
definite metric for the black hole interior.  

This argument generalizes to any situation in which part of the
spacetime remains inaccessible to a given observer and the metric
depends strongly on the outcome of quantum fluctuations.  By
definition, both of these conditions are satisfied in eternal
inflation.  For example, an observer cannot predict when and where
bubbles of new vacua will form.  He would need to describe the
inaccessible regions in terms of a superposition of all possible decay
sequences.  Because the formation of a bubble affects the large-scale
geometry, this task transcends the realm of semiclassical gravity.
Therefore, a local observer in eternal inflation cannot consistently
define a global geometry.

We conclude that the local and the global point of view are really
inequivalent.  This is encouraging since it opens the possibility for
a probability measure to arise naturally from the local point of view,
without the ambiguities that plague the global
description~\cite{Bou06}.  (If eternal inflation admits no global,
classical geometry, then the problem is not which slicing to pick.
The problem is that there is nothing to slice.)

The remainder of the paper is devoted to an analyzing how local
observers experience an eternally inflating universe.  An important
difference to the global view is that for any one observer, inflation
will eventually end (except if all vacua have positive energy---a
relatively simple case that is hard to reconcile with
observation~\cite{DysKle02}).  We focus not only on observations
during eternal inflation, but also consider the observer's fate in the
asymptotic future after inflation has ended.  We distinguish between
SREI (slow roll eternal inflation) and FVEI (false vacuum eternal
inflation).

In Sec.~\ref{sec-SREI}, we review the conventional global description
of SREI.  We then offer a local description.  A typical worldline in
SREI experiences nothing more than ordinary slow-roll.  In particular,
the entropy will not decrease by more than 1 as a result of
fluctuations, as required by the second law of thermodynamics.

Along any given geodesic, SREI eventually must end. If the effective
cosmological constant becomes negative, there will be a big crunch,
long before the scales that left the horizon during SREI would
re-enter.  In regions with positive cosmological constant, the same
censorship is achieved by a de~Sitter horizon.\footnote{This is true
  even if the vacuum eventually decays, as long as it lives long
  compared to the Hubble time \cite{KalKle04}.}

The case of zero cosmological constant is more subtle, and we treat it
in more detail.  Curvature perturbations become of order one precisely
when scales originating in the era of SREI re-enter.  Such large
perturbations rapidly cause gravitational collapse, trapping the
observer inside a black hole.  Thus, typical observers cannot probe
the scales produced during SREI.

In Sec.~\ref{sec-fvei}, we describe false vacuum eternal inflation
(FVEI), in which the inflaton is stuck in a metastable minimum and
must tunnel to exit inflation. FVEI shares many qualitative features
with SREI. If several metastable vacua are available, then a typical
worldline tunnels repeatedly, lowering the vacuum energy until it
reaches a terminal vacuum. As in the slow-roll case, some highly
atypical worldlines take a detour up to higher vacuum energy on the
way to reaching a terminal vacuum.

The asymptotic structure of regions of vanishing vacuum energy differs
strongly from SREI.  In the leading approximation, FVEI produces an
open FRW universe in which perturbations stop growing after curvature
dominates.  A more careful treatment should take into account
collisions with other bubbles of true vacuum~\cite{GutWei83}.  We
construct an exact solution describing the collision of two zero
cosmological constant bubbles in an ambient de Sitter space.  We
assume that the energy of the colliding walls is converted to dust.
Even in this worst case scenario, we find that bubble collisions do
not cause gravitational collapse.  Hence, the asymptotic conformal
structure is that of an unperturbed open FRW universe (the hats shown
in Fig.~\ref{fig-fveipenrose}).  Asymptotic observers do not fall into
black holes, and can look back on the infinite cluster of bubbles
colliding with their own.

The appendix summarizes various results on slow roll inflation used in
the main text.

\section{Predicting the geometry behind the horizon}
\label{sec-gedanken}

The macroscopic geometry inside a black hole can usually be predicted
by an external observer, if initial conditions are known.  In this
section we show that this becomes impossible if the geometry depends
strongly on the outcome of quantum measurements in the invisible
region.  This implies that semiclassical gravity cannot describe the
region behind the horizon of a local observer in eternal inflation.

\subsection{A gedankenexperiment involving a black hole}
\label{sec-ged1}

\begin{figure}
\begin{center}
\includegraphics[scale = .5]{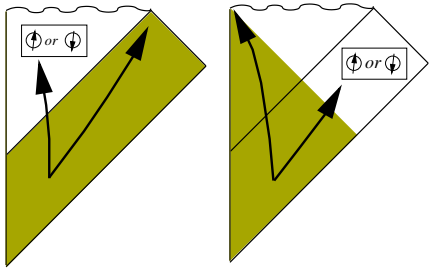}
\end{center}
\caption{On the left, a spin is measured after the apparatus has
  entered a black hole, leaving the outside observer ignorant of the
  outcome.  On the right, the measurement takes place outside the
  black hole, beyond the causal past of an infalling observer.  Either
  way, the observer cannot ascribe a definite outcome to the
  measurement.  If the geometry depends on this outcome, it cannot be
  treated semiclassically, since it would involve a quantum
  superposition of macroscopically distinct metrics.}
\label{fig-bhexp}
\end{figure}

Consider the gedankenexperiment shown in Fig.~\ref{fig-bhexp}.
An electron is prepared in a linear superposition of spin-up and
spin-down, $|\psi\rangle = \alpha\, |\!\!\!\uparrow\rangle + \beta\,
|\!\!\!\downarrow\rangle$, and placed inside an apparatus.  Then the
entire apparatus, including the electron, is thrown into a very large
black hole.  The apparatus is programmed to measure the spin only
after crossing the event horizon of the black hole, but well before it
reaches regions whose curvature is important on the scale of the
apparatus.

Because the apparatus is freely falling, and because curvature is
negligible, the measurement of the spin proceeds exactly as it would
in flat space.  The apparatus couples to the spin, resulting in the
joint state
\begin{equation}
  |\Psi\rangle = \alpha \,|\!\uparrow\rangle_A \otimes 
  \,|\!\uparrow\rangle +
  \beta \,|\!\downarrow\rangle_A \otimes \,|\!\downarrow\rangle~.
\end{equation}
The subscript $A$ refers to the state vector in the Hilbert space of
the apparatus.  Of course, a superposition of macroscopically
different states, such as the apparatus pointing up and pointing down,
is never observed, and we understand why.  A macroscopic object cannot
be insulated from the environment; for example, it will emit infrared
photons.  

Thus, a full description must include the environment.  What is
perceived as the apparatus will really be entangled with many external
degrees of freedom that cannot be observed and thus are implicitly
traced over.  One can show~\cite{Schl03} that this results
in a density matrix describing the apparatus-electron part of the
world, with probability $|\alpha|^2$ for the state
$|\!\uparrow\rangle_A \otimes \,|\!\uparrow\rangle$ and probability
$|\beta|^2$ for the state $|\!\downarrow\rangle_A \otimes
\,|\!\downarrow\rangle$. The description of the apparatus alone thus
exhibits {\em decoherence\/}, which allows us to understand the
apparent nonunitarity of a definite measurement outcome in terms of a
lack of information about inaccessible degrees of freedom.

However, the description from the point of view of an observer outside
the black hole is very different.  From her point of view, the
information contained in the infalling apparatus is imprinted on the
black hole horizon.  After the black hole evaporates, the information
will be encoded in Hawking radiation.  (At least, the weight of the
evidence from string theory suggests that this is a unitary process.)
Thus, while the outside observer can predict that the apparatus will
measure the electron, she cannot ever state that the electron was
measured to be up (or down).  Because the apparatus becomes correlated
with the electron, she would ascribe to the black hole interior a
superposition of macroscopically distinguishable states.

This can be understood also as follows.  From the outside point of
view, the black hole includes all the degrees of freedom that the
apparatus can interact with after measuring the electron.  That is, it
includes the whole environment accessible to the apparatus.  With
these degrees of freedom included, the evolution should indeed be
unitary.  Thus, a natural\footnote{It may be possible to identify an
  extremely complicated factorization of the Hilbert space of the
  Hawking radiation that is dual to the factorization into the
  apparatus and environment Hilbert spaces, and to consider a
  (practically inconceivable) experiment whereby only the correlations
  in the Hawking radiation corresponding to the ``apparatus'' are
  measured.  The resulting density matrix would then be dual to the
  one inside the black hole.  But we see no reason to expect, nor
  means to verify, that the outcomes (spin up or spin down) of the two
  experiments would agree.} extrapolation of the outside data implies
that the electron, apparatus, and the rest of the black hole interior
remain in a superposition of ``up'' and ``down''.

Now suppose that the apparatus involves very massive moving parts.
For example, its pointer may be a neutron star, which is moved to
different places depending on whether the electron spin is up or down.
The point is to arrange for the gravitational field inside the black
hole to depend strongly, and macroscopically, on the outcome of the
measurement.  We concluded earlier that from the outside point of
view, the black hole interior must be described by a superposition of
macroscopically distinguishable states.  Now we see that this may
involve a superposition of very different metrics.  Therefore, there
is no unique semiclassical geometry that can be predicted for the
black hole interior by an exterior observer, even if the initial
quantum state is known exactly.

Normally, two different observers would agree on the outcome of a
quantum-mechanical measurement.  What distinguishes the above
gedankenexperiment is that the location of the experiment, the black
hole interior, never enters the past of the outside observer.
Therefore the outside observer (and more importantly, environmental
degrees of freedom she interacts with) never get to look at the
apparatus after the measurement.  It is not particularly important
that the causal separation happens to manifest itself in a black hole
event horizon.

In fact, we can easily switch roles.  Suppose now that the apparatus
containing the electron remains far outside the black hole, and is
programmed to measure the spin of the electron at a time $t_0$.
Consider an observer who falls into the black hole and ends up at the
singularity, as shown in Fig.~\ref{fig-bhexp}.  This observer's
causal past does not include all of the black hole exterior.  If $t_0$
is chosen sufficiently late, then the infalling observer cannot
witness the decoherence of the apparatus, and so he cannot predict a
unique semiclassical geometry for the outside region.

This illustrates that the relevant causal boundary is simply the
boundary of the past of the observer's worldline.  In the first
experiment, this happens to be a horizon. In the second, it is the
past light-cone of the observer at the moment when he hits the
singularity.  The only difference is that in the second case, there
are additional uncertainties because of unknown initial conditions at
a portion of past null infinity.  We will not be interested here in
this trivial obstacle to predictability; where necessary, we will
assume that initial conditions are completely known.

\begin{figure}
\begin{center}
\includegraphics[scale = .5]{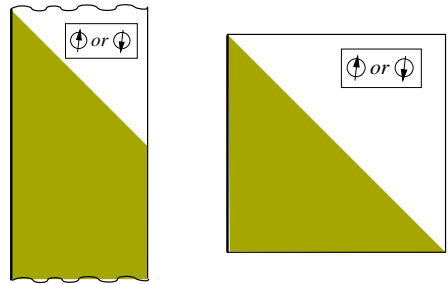}
\end{center}
\caption{Examples of causal horizons in cosmology: A closed FRW
  universe (left) and de~Sitter space (right).  The observer is on the
  left edge of each diagram.  She cannot access the result of the
  measurement in the (white) region outside her causal past.}
\label{fig-cosexp}
\end{figure} 

Therefore, our reasoning generalizes to other nontrivial spacetimes,
and in particular, to cosmology.  Examples are shown in
Fig.~\ref{fig-cosexp}: an observer hitting a big crunch, and an
observer surrounded by a cosmological horizon in de~Sitter space.

Let us summarize.  The emergence of classical behavior, including
classical geometry, depends on the separation of degrees of freedom
into gross, observed features and unobserved ``environmental'' degrees
of freedom.  The selection of a definite outcome from the
possibilities allowed by the resulting density matrix cannot be
unitarily predicted from initial data.  But the metric generally
depends on this outcome.  {\em Therefore, an observer cannot ascribe a
  unique geometry to regions that are forever outside his past, even
  if initial conditions are known and even if another observer,
  entering such a region, can be sure to find a well-defined
  geometry.}

\subsection{Eternal inflation}
\label{sec-ged2}

Now let us apply this result to eternal inflation.  In FVEI, an
infinite number of bubbles are nucleated globally, but most will 
be outside the observer's horizon.  Hence, to the
observer, they cannot be ascribed a definite place and time.  From a
field theory point of view, the region beyond the horizon would need
to be described by a wave function with slowly increasing support in
the lower energy vacua.  When gravity is included, this corresponds to
a superposition of different geometries.

SREI relies on the thermal fluctuations of the inflaton field.  We
assume that inflation eventually ends for any observer, so most of the
SREI regime is outside the observer's horizon.  The fluctuations in
this region do not decohere.  (In fact, the results of the following
two sections show that that they are unlikely to decohere even inside
the horizon.)  In particular, it is impossible to identify specific
spacetime regions where the unlikely repeated upward fluctuations
occur that drive SREI.  From a field theory point of view, the
wavefunction for the inflaton field simply spreads out outside the
horizon.  In most places the spread is not large enough to dominate
over the classical evolution, and an approximate geometry can be
assigned.  But SREI is eternal only due to the tails of this
wavefunction.  In the usual description, these tails set the classical
value of the field in a small fraction of the spacetime volume.  The
single observer cannot know where, and thus cannot assign a definite
geometry.

Note that we did not assume an ignorance of initial conditions.  If
one imagines that the universe was created, perhaps by ``tunneling
from nothing'', there might be an initial hypersurface, and
generically, an observer's past lightcone will only include a fraction
of that surface.  However, for the purposes of our argument we can
happily assume that the observer knows the entire initial data,
perhaps because they are fixed by some physical law (e.g., the
no-boundary proposal).  

The point is that a definite geometry is incompatible with the unitary
quantum evolution of any initial data that lead to eternal inflation.
A semiclassical description of the geometry depends on the decoherence
of quantum effects such as bubble formation and field fluctuations.
The geometry in our past is classical precisely because of
nonunitarity: it has decohered to one shape and not another.  But from
a local observer's point of view, the regions beyond the horizon
remain in quantum superposition.

\section{Slow-roll eternal inflation}
\label{sec-SREI}

We will begin this section with a summary of the conventional global
description of slow-roll eternal inflation
(SREI)~\cite{Lin86b,GonLin87}.  We will then describe the experience
of a typical local observer, who simply experiences conventional
slow-roll inflation.  He will end up in the vacuum that lies along the
gradient from his starting point.  We shall see that this behavior is
in fact required by the second law of thermodynamics.  We will show
that at late times, all geodesics end on a singularity.

The enormous differences between the global and local description
should not come as a surprise.  SREI relies on unusually large or
persistent upward fluctuations of the inflaton field.  The global
picture compensates for the extreme rarity of such events by rewarding
them with large volume expansion factors.  But from the local point of
view, such fluctuations are simply extremely unlikely, with no
compensating features.  For the reasons given in
Sec.~\ref{sec-gedanken}, 
volume factors in
unobservable regions are meaningless.

\subsection{Conventional global description}
\label{sec-conv}

The conventional argument for slow-roll eternal inflation goes as
follows.  Consider one Hubble volume during slow-roll inflation.
After one Hubble time, it will have expanded to $e^3\approx 20$ new
Hubble volumes.  Meanwhile, the expectation value of the inflaton
field will have decreased by an amount
\begin{equation}
  \Delta\phi = \frac{|\dot{\phi}|}{H} 
  = \frac{V'}{3H^2} = \frac{H'}{4\pi H}
\end{equation}
from rolling classically down the potential $V(\phi)$.

By Eq.~(\ref{eq-pphi2}), the quantum fluctuations of $\phi$ on the
Hubble scale are given by
\begin{equation}
  {\cal P}_\phi^{1/2}(H) = \frac{H}{2\pi}~.
\end{equation}
A measurement of the inflaton field should yield values in a Gaussian
distribution about the classical value of $\phi$, with standard
deviation $H/2\pi$.  The Hubble scale quantum fluctuations will
dominate over the classical evolution if
\begin{equation}
\frac{H^2}{H'}\gtrsim  1~.
\label{eq-etcond}
\end{equation}

Let us suppose that this condition is comfortably satisfied:
$H^2/H'\gg 1$.  Then every Hubble time, approximately half of the 20
new Hubble volumes should witness an increase in the inflaton field
value.  Instead of rolling down, these ten Hubble volumes will have
climbed up the inflationary potential.

Of course, the other ten Hubble volumes will see a decrease of $\phi$,
mostly due to downward quantum fluctuations rather than classical
slow-roll.  Over time, some regions evolve into a regime where
$H^2/H'<1$.  Thus, regions are produced where classical evolution
dominates, and inflation can come to an end.

To get eternal inflation, it is enough that $\phi$ increases in
at least one new Hubble volume every Hubble time.  For this, it
suffices to demand $H^2/H'\gtrsim 1$.  Then there will always be some
regions where inflation continues.  Inflation is globally eternal and
ends only locally.

Let us assume that the inflaton potential vanishes at the global
minimum $\phi=0$ and contains a quantum-dominated regime:
\begin{equation}
H^2/H'>1~~~~\mbox{for all}~\phi>\phi_{\rm qc}~.
\end{equation}
We will refer to $\phi_{\rm qc}$, and the era when this value is
attained, as the {\em quantum/classical boundary}.  For definiteness,
we take $\phi_{\rm qc}>0$ to be positive.  We also assume that $V\ll
1$ in the quantum-dominated regime.  It should be stressed that SREI
does not require Planck scale densities or curvatures.

\begin{figure}
\begin{center}
\includegraphics[scale = .5]{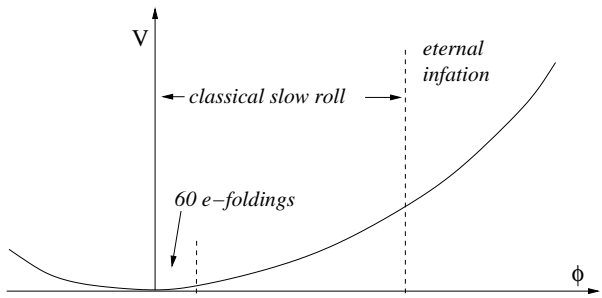}
\end{center}
\caption{The slow-roll potential $V=\frac{1}{2}m^2\phi^2$.  The SREI
  regime is $\phi>m^{-1/2}$.}
\label{fig-quadpot}
\end{figure}

Not all sensible inflationary potentials contain such a regime.  But
it is fairly generic, in that it requires only that the potential
contain portions that are sufficiently high (for a large effective
cosmological constant to drive the quantum fluctuations) and not too
steep (so that the classical slow-roll is negligible).  For example,
let us suppose that the inflaton potential was $V=m^2 \phi^2/2$, with
no corrections even for large $\phi$, and with $m\sim O(10^{-6})$.
Then the quantum-dominated regime corresponds to $\phi\gtrsim
m^{-1/2}$ (see Fig.~\ref{fig-quadpot}).

In this example, the inflaton would have been at the classical/quantum
boundary, $\phi_{\rm qc}\sim m^{-1/2}$, at a very early time during
inflation, namely $O(m^{-1})$ e-foldings before entering the
range, $\phi\sim O(1)$, that created the presently visible universe.
Correspondingly, we would have to wait for the universe to expand by
an enormous factor (in this example, $\exp[O(m^{-1})]$) before we
could see scales that left the horizon near the eternally inflating
regime.  The corresponding comoving wavenumber, $k_{\rm qc}$,
satisfies
\begin{equation}
\log\frac{k_{\rm qc}}{k_0} \sim O(m^{-1})~.
\end{equation}

\begin{figure}
\begin{center}
\includegraphics[scale = .5]{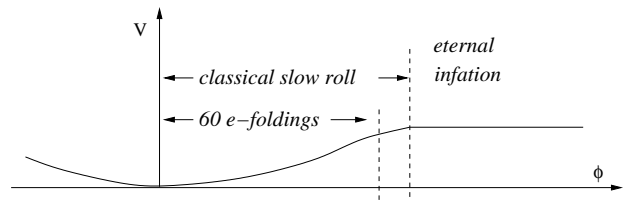}
\end{center}
\caption{A potential with a plateau supporting eternal inflation and
  a relatively short-lived classical slow roll.}
\label{fig-mesa}
\end{figure}

Thus, the scale associated with the quantum/classical boundary tends
to be exponentially large compared to the scale $k_0$ that is
presently entering the horizon.  However, one could tune the inflaton
potential to have a smaller separation between the observable scale
and the scale corresponding to the quantum/classical boundary (see
Fig.~\ref{fig-mesa}).  Our conclusions will be independent of the size
of this hierarchy.

\subsection{Horizon area and second law}
\label{sec-local}

During the inflationary era, the local description of SREI is quite
similar to that of ordinary slow roll inflation:  The field rolls down
classically, following the gradient of the potential.  Typical quantum
fluctuations are eventually overcome by classical evolution.

To see this, recall that the quantum fluctuations (the spread of the
wavefunction) of $\phi$ grow like
\begin{equation}
\delta_{\rm q}\phi(t)\approx {H \over 2 \pi}  \sqrt{Ht}~,
\label{eq-poui}
\end{equation}
where we assume $Ht>1$; see Eq.~(\ref{eq-randomwalk}).  
Meanwhile, the classical slow roll causes $\phi$ to
decrease by
\begin{equation}
\Delta\phi(t) \approx |\dot{\phi}| t = \frac{V'}{3H} t =
\frac{H'}{4\pi} t~.
\label{eq-cldelphi}
\end{equation}
The above equations hold both in SREI and in ordinary slow-roll
inflation.

In SREI, the classical evolution begins to dominate after a time
\begin{equation}
H t_{\rm c} = \left(2 \frac{H^2}{H'}\right)^2~.
\label{eq-dom}
\end{equation}
Recall that $H^2/H'>1$ in SREI, so quantum fluctuations dominate for
longer than one Hubble time. During this time the field changes classically by
$H^3/(\pi H')$.

But classical evolution eventually wins out, and one can predict with
great likelihood which vacuum it will lead to.  The only case in which
there is significant uncertainty of the outcome is if the field starts
out within $H^3/H'$ of a local maximum of the potential.  In this case
the observer cannot tell which side she is on.

Let us identify and resolve a possible paradox.  The
Bekenstein-Hawking entropy of the cosmological horizon,
\begin{equation}
S = \frac{A}{4} = \frac{3}{8V(\phi)},
\end{equation}
depends on the effective cosmological constant $\Lambda = 8\pi
V(\phi)/3$, and thus on $\phi$.  During classical slow-roll, $\Lambda$
will decrease, and the entropy will increase.

However, during SREI the quantum fluctuations dominate, and the area
is about as likely to decrease as it is to increase.  It might appear,
therefore, that SREI conflicts with the second law of thermodynamics.
Of course, the second law will occasionally be violated in any system,
but such fluctuations must be exponentially rare.  In SREI, however,
area-decreasing fluctuations occur about half of the time.  We will
now show that SREI is nevertheless consistent with the second law.

The change in horizon area, $\delta_{\rm q}A$, induced by quantum
fluctuations of the inflaton field in the time $t$, can be computed
from Eq.~(\ref{eq-poui}).  Since $A\sim H^{-2}$ one has
\begin{equation}
\delta_{\rm q}A(t)\sim \frac{H'}{H^3} \delta_{\rm q}\phi 
\sim \frac{H'}{H^2} \sqrt{Ht}~.
\label{eq-dqa}
\end{equation}
Until the time
\begin{equation}
\sqrt{Ht} \sim \frac{H^2}{H'}~,
\label{eq-htqa}
\end{equation}
this fluctuation is less than 1, corresponding to a negligible
decrease in entropy.  In particular, it follows that the quantum
fluctuation during one Hubble time, $\delta_{\rm q}\phi\sim H$, causes
an undetectable change in horizon area and entropy, as long as the
SREI condition, Eq.~(\ref{eq-etcond}), is satisfied.

One might be concerned that the second law does become violated after
the time given in Eq.~(\ref{eq-htqa}).  But all is well, for it is
precisely after this time that the classical evolution overwhelms the
quantum effects, by Eq.~(\ref{eq-dom}).  

In fact one can see this more directly.  The classical increase in
horizon area is given by
\begin{equation}
\Delta A(t) \sim [\delta_{\rm q}A(t)]^2~. 
\end{equation}
according to Eqs.~(\ref{eq-cldelphi}) and (\ref{eq-dqa}).  It is always
positive, corresponding to an increase in entropy.  It dominates
($\Delta A>\delta_{\rm q}A$) as soon as either area change becomes
larger than the Planck area ($\delta_{\rm q}A>1$).

\subsection{The fate of a generic observer}
\label{sec-fate}

How about the experience of late-time postinflationary observers?  We
will now show that generically, their past light-cones will not grow
large enough to include modes that were produced during the eternal
phase.  As discussed in the introduction, this is obvious if the
cosmological constant is nonzero, so we will focus on the $\Lambda=0$
case.

The curvature perturbation on the comoving scale $k^{-1}$ is given by
Eq.~(\ref{eq-curvpert}):
\begin{equation}
  {\cal P}_{\cal R}^{1/2}(k) = \frac{H_*^2}{H_*'}~.
\label{eq-curvpert2}
\end{equation}
But according to Eq.~(\ref{eq-etcond}), the quantum-dominated regime
satisfies $H^2/H'>1$.  Hence, the curvature perturbations produced
during this regime are of order one, and perturbation theory breaks
down.  

What would this look like from the point of view of a
post-inflationary observer?  The perturbations we see today were
produced towards the end of inflation, when $H^2/H'$ was small (${\cal
  P}_{\cal R}\sim 10^{-5}$).  So first we would have to wait for an
exponentially long time compared to the current age of the universe.
Then we would begin to see much larger scales that left the horizon
earlier in inflation, when $H^2/H'$ was larger.

As we approach the time when the comoving scale $k_{\rm qc}$ enters
the horizon, the spatial slices will no longer look flat but will be
noticably closed or open:
\begin{equation}
\Omega-1 \sim O(1)~,
\end{equation}
where $\Omega=\rho/\rho_{\rm c}$ and $\rho_{\rm c}=3H^2/8\pi$ is the
critical density.  Whether it looks closed or open depends on the sign
of the perturbations that most recently entered the horizon; this will
itself fluctuate.

When the perturbations become of order unity, overdense fluctuations
will close the universe and cause gravitational collapse on a distance
scale comparable to the horizon.  The timescale for this collapse is
also comparable to the horizon scale (i.e., the age of the universe).
Thus, the observer ends up in a crunch, and cannot see further
than the comoving distance $k_{\rm qc}^{-1}$.

Let us make this more precise.  A region decouples from the
cosmological flow and the perturbation becomes nonlinear when the
density contrast, $\delta\equiv\delta\rho/\rho$, becomes of order one.
The dust then stops expanding and begins to recollapse.  One can show
that the ``turnaround'' (when the local expansion rate momentarily
vanishes) occurs roughly at the time $t_{\rm turn}$ when
\begin{equation}
{\cal P}_\delta(k)\approx 1~,
\end{equation}
as computed from the linear theory.  By Eqs.~(\ref{eq-lncvpb}) and
(\ref{eq-lnmtpb}),
\begin{equation}
  {\cal P}_\delta = \left(\frac{k}{aH}\right)^4 {\cal P}_{\cal R}~.
\end{equation}
Hence,
\begin{equation}
H(t_{\rm turn})\,a(t_{\rm turn}) = k\, {\cal P}_{\cal R}^{1/4}(k)~,
\end{equation}
where the right hand side is independent of time.  

In the matter-dominated phase, $Ha = \dot{a} = \frac{2}{3} t^{-1/3}$.
Neglecting factors of order one, we find
\begin{equation}
t_{\rm turn} \sim 
\left[\frac{1}{k\, {\cal P}_{\cal R}^{1/4}(k)}\right]^3~.
\end{equation}
In conformal time, $\eta = \int dt/a = 3 t^{1/3}$, the turnaround
occurs at
\begin{equation}
\eta_{\rm turn} \sim  \frac{1}{k\, {\cal P}_{\cal R}^{1/4}(k)}~.
\end{equation}

\begin{figure}
\begin{center}
\includegraphics[scale = .5]{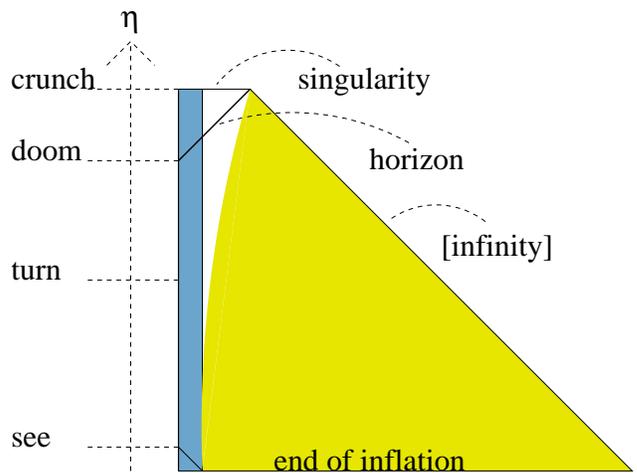}
\end{center}
\caption{Even a weak perturbation will eventually go nonlinear, but
  there is plenty of time to avoid getting trapped in the collapse.
  This conformal diagram shows an exact solution constructed with the
  help of Birkhoff's theorem.  The blue region is a small portion of a
  closed universe (the polar cap of the three-sphere).  Because its
  curvature radius is much larger than its size, it represents only a
  slightly overdense region embedded in a flat universe.  The world
  volume of the unperturbed dust particles is shown in yellow.  The
  white region is vacuous and is a portion of the Schwarzschild
  solution.  It develops as the overdense region decouples from the
  ambient expansion.  The perturbation is first fully seen at
  $\eta_{\rm see}$.  It goes nonlinear much later, at $\eta_{\rm
    turn}$.  Only at $\eta_{\rm doom}$ does the inert observer enter a
  black hole.}
\label{fig-weak}
\end{figure}

Let us compare this to important earlier timescales (see
Fig.~\ref{fig-weak}).  The perturbation becomes first visible when
\begin{equation}
\eta_{\rm see} \approx \frac{1}{k}~.
\label{eq-etasee}
\end{equation}
This is the time at which an observer can see a comoving piece of size
$k^{-1}$ of the reheating, or CMB hypersurface.  (We approximate both
of these times as $\eta=0$.)  Since the (past) apparent horizon
satisfies $\eta=2 x$, the perturbation enters the horizon a short
while later, at $\eta=2/k$.  Note that if the perturbation is small,
${\cal P}_{\cal R}(k)\ll 1$, then the turnaround time is very late
compared to the time when the perturbation is first noticed.

We could assume more optimistically that the observer can detect the
fluctuation as soon as it freezes, i.e., when the scale $k$ at the
time of freezing enters the observer's past light-cone.  This would
yield a slightly earlier detection time
\begin{equation}
  \eta_{\rm see} \approx 
  \frac{1}{k}~\frac{N_{\rm freeze}}{N_{\rm freeze}+1}~,
\end{equation}
where $N_{\rm freeze}$ is the factor by which the mode expands between
horizon exit and freezing; see Eq.~(\ref{eq-freeze}) and
Fig.~\ref{fig-freeze}.  Since $N_{\rm freeze}$ is of order ten, this
is not a significant correction.  Moreover, it is not clear whether
signals are available that would make such an early detection possible
even in principle.  Hence, we will work with Eq.~(\ref{eq-etasee}).

Let us also consider important timescales after the turnaround.  Since
the overdense region behaves like a portion of a closed, matter
dominated FRW universe, a singularity occurs at the time $t_{\rm
  crunch} \approx 2 t_{\rm turn}$, or
\begin{equation}
  \eta_{\rm crunch} \approx 2 \eta_{\rm turn}  
  \sim  \frac{2}{k\, {\cal P}_{\cal R}^{1/4}(k)}~,
\end{equation}
using the conformal time appropriate for a closed
universe.  Here we assume that pressure is negligible during the
collapse, which is a good approximation for the very large scales we
are interested in.

Viewed from outside the overdense region, a black hole will appear to
form, and the singularity will be hidden behind an event horizon.  The
event horizon meets the (future) apparent horizon, $\eta_{\rm
  crunch}-\eta = 2x$, at the outer limit of the overdense region, at
$x=k^{-1}$.  Hence, an observer at $x=0$ crosses the event horizon at
\begin{equation}
  \eta_{\rm doom} = \eta_{\rm crunch}-\frac{3}{k}
  \sim \frac{1}{k}\left[ \frac{2}{{\cal P}_{\cal R}^{1/4}(k)}-3\right]~.
\end{equation}

The time between noticing an overdensity, and entering the black hole
it leads to, is therefore
\begin{equation}
\eta_{\rm doom} - \eta_{\rm see} \sim 
 \frac{1}{k}\left[ \frac{2}{{\cal P}_{\cal R}^{1/4}(k)}-4\right]~.
\end{equation}

For small perturbations (Fig.~\ref{fig-weak}), we thus see that the
small parameter ${\cal P}_{\cal R}$ leads to large separations between
$\eta_{\rm see}$, the time when the perturbation is first detected;
$\eta_{\rm turn}$, the time when it becomes nonlinear; and finally,
$\eta_{\rm doom}$, the time when black hole forms and the singularity
is near.  This leaves an observer plenty of time for escape from the
overdense region, by moving to an underdense region after detecting a
perturbation that would eventually enclose him in a black hole.

\begin{figure*}
\begin{center}
\includegraphics[scale = .5]{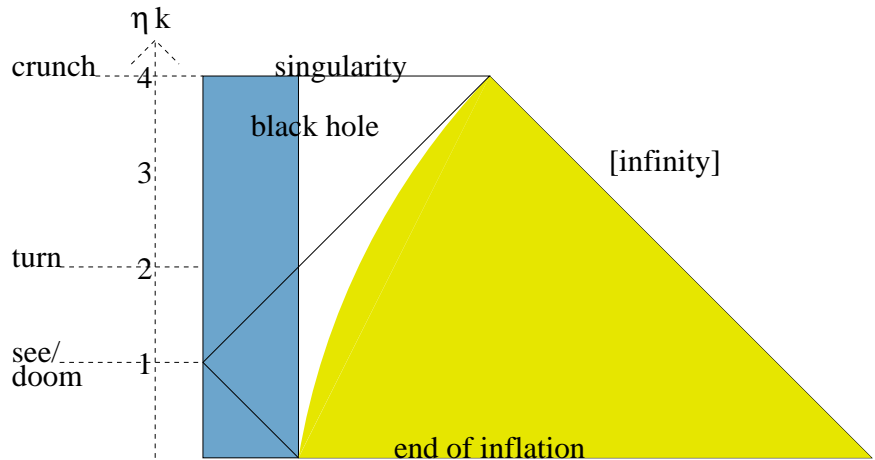}
\end{center}
\caption{A strong perturbation catches the observer unaware (compare
  to Fig.~1).  This conformal diagram shows another exact
  solution, but this time we keep half of the three-sphere of the
  closed universe (blue).  This represents a strong perturbation
  because its size is comparable to its curvature radius.  The
  observer can see the overdense region only as he is already falling
  into the black hole it forms.}
\label{fig-strong}
\end{figure*}

However, for larger perturbations, the separation of timescales
shrinks.  As longer wavelengths enter the horizon, there is less and less
time for the observer to react and move to safety.  Overdense
perturbations of order one are lethal: For ${\cal P}_{\cal R}\sim 1$,
\begin{equation}
\eta_{\rm doom}\lesssim\eta_{\rm see}~.
\end{equation}
By the time the observer can measure the curvature, he is already
engulfed in a black hole and doomed to crunch.\footnote{Recall that
  because of the nonlocality of event horizons, it is possible to
  enter a black hole in a locally innocuous environement.}  An example
is shown in Fig.~\ref{fig-strong}. 

Of course, not every perturbation is overdense; at any given time, it
is just as likely that the next fluctuation to become visible will be
underdense.  But from the perspective of a local observer, one
overdense perturbation is enough to cause a singularity.  As the
deformations of the metric due to inflaton field fluctuations become
larger, there will eventually be an overdense fluctuation that will
engulf the observer in a black hole.

\section{False vacuum eternal inflation}
\label{sec-fvei}

We will begin this section by reviewing the conventional global
description of FVEI.  Next, we will describe the experience of a
typical observer, and identify which aspects of the global description
are unobservable. We find that observers who exit FVEI into a zero
cosmological constant vacuum experience an open exapanding FRW
universe. The presence of asymptotic regions with $\Lambda=0$ sharply
distinguishes FVEI from SREI. We will argue that the inevitable
collisions between true vacuum bubbles do not lead to gravitational
collapse on large scales, so the asymptotic $\Lambda = 0$ regions
persist even after collisions are taken into account. 
We will support our claim by finding an
exact solution describing the collision of two true vacuum bubbles.
Furthermore, an
infinite number of bubble collisions are contained within the horizon
of an observer who exits to $\Lambda = 0$.

\subsection{Conventional global description}

The starting point for the global description of FVEI is the global de
Sitter geometry associated with a metastable vacuum.  Decay occurs by
bubble nucleation, as described by Coleman and de~Luccia
\cite{CDL}. Typically the new vacuum has lower cosmological constant
(otherwise, interestingly, it cannot really be represented globally as
a nucleation event in a background geometry). The bubble expands into
the ambient de Sitter space, asymptotically approaching the speed of
light.  However, because the de Sitter space is itself expanding, it
is not completely eaten by the bubble. 

Eventually an infinite number of bubbles will nucleate, eating all of
the comoving volume.  However, the proper volume of each of the false
vacua continues to grow.  Vacua with $\Lambda \leq 0$ do not decay
further and are called terminal. As described by Guth and Weinberg
\cite{GutWei83}, a given bubble of true vacuum eventually collides
with an infinite number of other bubbles. However, in the regime where
inflation is eternal, the true vacuum regions form disconnected islands in the sea
of false vacuum. Each island consists of an infinite number of
bubbles, and there are an infinite number of islands.

A Penrose diagram of the global geometry is shown in
Fig.~\ref{fig-fveipenrose}.  As discussed in Sec.~\ref{sec-gedanken},
this geometry cannot be constructed from initial conditions without
appealing to an infinite number of causally disconnected observers.

\subsection{The fate of a generic observer}

Just like in SREI, for any given worldline, inflation will not be
eternal.  The worldline will simply experience a sequence of tunneling
events (typically decreasing the cosmological constant).  If there are
terminal vacua, it will end up in one after a finite time.

Thus, no single observer can see the complicated fractal global
geometry described above.  Observers who end up in a region with
negative cosmological constant experience a big crunch within a time
set by the curvature of the AdS space (see for example \cite{BJ}).

Observers in $\Lambda=0$ regions are much luckier.  As the solution
below shows, they find themselves in an expanding open FRW
universe. Depending on the details, gravitational collapse can occur
on short length scales. However, on sufficiently large length scales
the curvature dominates and collapse cannot occur.

We describe the solution in some detail because we will need it in
order to analyze bubble collisions in the following section. 
We confine ourselves
to the thin wall approximation for simplicity, but indicate 
which features persist beyond
this approximation.

\begin{figure}[!htb]
\center
\includegraphics [scale=.5] {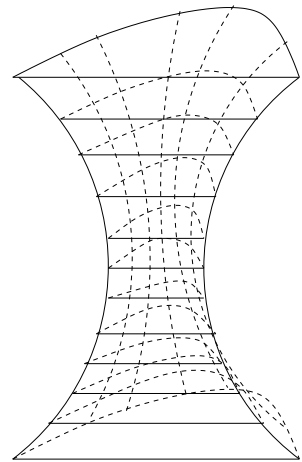} \caption{\small\sl The
  Coleman-De~Luccia bounce geometry, as an embedding in Minkowski
  space.  The flat piece has zero cosmological constant, while the
  curved piece has positive cosmological constant.  A domain wall
  separates the two regions.} \label{embd}
\end{figure}

The single bubble solution consists of a region of true vacuum
surrounded by a domain wall, with false vacuum outside.  Inside the
domain wall, the spacetime is approximately empty, so the geometry is
simply a region of Minkowski space glued across the thin domain wall
to a portion of de Sitter space.  The domain wall is a hyperboloid,
whose worldvolume describes a three-dimensional de Sitter spacetime.
Its curvature radius is $r_0$, where $r_0$ is
determined by the tension $\sigma$ of the domain wall and the
($4$-dimensional) de Sitter radius $R$,
\begin{equation} 
  r_0 = {8 \pi  \sigma R^2 \over (4 \pi 
    \sigma R)^2 + 1 }~.  
\end{equation}
In the flat spacetime enclosed by the hyperboloid, the position of the
domain wall satisfies
\begin{equation}
  X_\mu X^\mu = -r_0^2~, 
\end{equation}
where $X_\mu$ are the standard Minkowski coordinates.  The wall
expands from a minimum radius $r_0$, at $X_0=0$, to infinite size as
$X_0 \to \pm \infty$.  To describe the motion of the domain wall from
the de Sitter side, it is simplest to embed de Sitter space in
$5$-dimensional Minkowski space as the surface
\begin{equation}
 X_\mu X^\mu = -R^2, ~~~~~\mu = 0,1,...,4 
\end{equation}
The domain wall is the intersection of the de Sitter space with
the plane
\begin{equation}
 X_4 = (R^2 - r_0^2)^{1/2}. 
\end{equation}
The part of the space farther from the origin than the plane is
discarded and replaced by the piece of flat space described above. An embedding
picture of the solution is shown in figure \ref{embd}.
The solution is invariant under the symmetry group of the hyperboloid,
$SO(1, 3)$. On the de Sitter side, the full $SO(1, 4)$ de Sitter
symmetry is broken down to those boosts and rotations which leave
$X_4$ unchanged. On the Minkowski side, the symmetry arises because
the domain wall picks out an origin, breaking translation invariance
while preserving the Lorentz group. 
The surviving $SO(1, 3)$ symmetry is not an artifact of the thin wall
approximation.
It is present in the full solution. 

One can write the metric in the flat space region in coordinates
adapted to the symmetry,
\begin{equation}
ds^2 = -dt^2 + a(t)^2 dH_3^2
\end{equation}
In the thin wall approximation, $a(t) = t$ and these coordinates
simply cover a piece of Minkowski space with hyperbolic time
slices. Beyond the thin wall approximation, the $SO(1,3)$ symmetry
which acts on $H_3$ is still present, but the scale factor
changes. Beyond the thin wall approximation, a uniform energy density
is present on the spatial slices, so the spacetime is an open expanding
FRW universe.

\subsection{Bubble collisions}

This picture is complicated by collisions with other true
vacuum bubbles.  However, we will now argue that they do not alter the
conclusion that gravitational collapse stops after curvature
domination.  For this purpose, we will present an exact solution
describing the collision of two bubbles of the same $\Lambda=0$
vacuum.

We will assume that all of the energy in the domain walls is converted
to dust upon colliding.  Since dust is generally more eager to
collapse than any other type of matter, we expect that our conclusion
extends to more general collisions.

We can use the de Sitter symmetries to make both bubbles time
symmetric with respect to the global de Sitter time. Then in the
embedding picture, each bubble is characterized by picking out a plane
which is parallel to the $X_0$ axis. (In the one-bubble description
above, we chose the plane $X_4 =\sqrt{R^2 - r_0^2}$.) The whole
embedding picture is then the de~Sitter hyperboloid cut by two
planes. At $X_0 = 0$, the two bubbles are separated by a region of
false vacuum, but as they grow they will collide. The surface along
which they collide is a $2$-dimensional surface defined by the
intersection of the de Sitter hyperboloid with the two planes: a
spacelike 2-hyperboloid.  When the true vacuum bubbles collide, the
energy in the domain walls is converted to dust, and we will need a
new solution to describe this part of the evolution.  The two-bubble
geometry preserves $SO(1,2)$ symmetry out of the original $SO(1,4)$
symmetry, because picking out the two planes is picking out two
special directions. Rotations and boosts which leave the two special
directions invariant are still symmetries of the problem.

We now construct a solution with the appropriate $SO(1,2)$ symmetry
which consists of two flat spaces connected across a wall of
dust. Since in the full solution dust is formed when the domain walls
annihilate, we will need to use only a piece of this solution. The
solution will be similar to the more familiar VIS solution in that it
will consist of two spatially compact (in the Minkowski time slicing)
regions of flat space joined across a wall of dust. Everything expands
as time goes on.

We choose the $SO(1,2)$ symmetry to act on the coordinates $X_0, X_1,
X_2$.  To preserve the symmetry, the solution should only depend on
the invariant interval $X_0^2 - X_1^2 - X_2^2$, and not on the
coordinates separately.  Defining a coordinate $\tau$ to be the proper
time along the worldline of a dust particle, the wall will be
described by two functions
\begin{align}
X_0^2 - X_1^2 - X_2^2 &= f_1(\tau) \\
X_3 &= f_2(\tau)
\end{align}
The motion of the wall is determined by the Israel junction
condition,
\begin{equation}
K_{ab} - h_{ab} K = 8 \pi T_{ab}
\end{equation}
where $K_{ab}$ is the extrinsic curvature of the wall, $h_{ab}$ is the
induced metric on the wall, $K$ is the trace of $K_{ab}$, and
$T_{ab}$ is the stress tensor of the wall, which in our case should
have a form appropriate for dust.

The solution is 
\begin{eqnarray}
X_0^2 - X_1^2 - X_2^2 &=& \tau^2 - L^2 \\
X_3 &=& b \pm L \cosh^{-1}({\tau \over L}).
\end{eqnarray}
Here $b$ is an arbitrary shift in the choice of origin and $L$ is a parameter
which depends on the energy of the colliding domain walls.
We can eliminate $\tau$ to find a single equation for the motion of the dust,
\begin{equation}
X_0^2 - X_1^2 - X_2^2 =  L^2 \sinh^2 \left({X_3  - b \over L}\right)
\label{eq-dustsoln}
\end{equation}
The intrinsic geometry of the dust is given by
\begin{equation}
ds^2 = -d\tau^2 + (\tau^2 - L^2) dH_2^2
\label{eq-intr}
\end{equation}
It is straightforward but tedious to check that this solution indeed
satisfies the appropriate junction conditions for dust. The dust
solution has a singularity at $\tau = L$; since we will only use part
of the dust solution, this singularity will not concern us.

Now we need to patch the dust solution onto the domain wall solution
to get our full solution. As noted above, the two true vacuum bubbles
collide on a spacelike 2-hyperboloid, so it is along such a
hyperboloid that we should match onto the dust solution. In the dust
solution, the plane $X_3=constant$ intersects the wall of dust in a
2-hyperboloid.

To match, we require the radius of curvature of the $2$-hyperboloid to
be equal to the radius of curvature $R_H$ along which the domain walls
collide,
\begin{equation}
\tau_0^2 - L^2 = R_H^2
\end{equation}
We also need energy conservation in the conversion of domain walls to
dust. The junction condition relates the energy density to the
parameters in the metric.  To match, at the time the dust is created
its energy should equal the center of mass energy $\rho_0$ of the
domain wall collision,
\begin{equation}
\rho_0 = {L \over 2 \pi (\tau_0^2 - L^2)}
\end{equation}
Combining these equations, we solve for the parameters of the dust
solution in terms of the parameters of the collision,
\begin{eqnarray}
 L &=& 2 \pi R_H^2 \rho_0 \\
\tau_0 &=& R_H \sqrt{1 + (2 \pi  R_H \rho_0)^2}
\label{eq-params}
\end{eqnarray}

Let us summarize the entire solution, focusing on an observer at the
center of one of the true vacuum bubbles. If there were just one true
vacuum bubble, such an observer would see true vacuum out to a domain
wall described by
\begin{equation}
  X_\mu X^\mu = -r_0^2~, 
\end{equation}
and false vacuum on the other side of the domain wall. Adding another
bubble means that we replace part of the domain wall with dust. We can
choose coordinates so that the part of the domain wall replaced by
dust is the region with $X_3 > a$. In the region it exists, the dust
is described by Eq.~(\ref{eq-dustsoln}).  Given a collision which
occurs along a $2$-hyperboloid with curvature radius $R_H$ and with
center of mass energy $\rho_0$, the parameters in the solution are
\begin{eqnarray}
 L &=& 2 \pi  R_H^2 \rho_0 \\
a &=& \sqrt{r_0^2 + R_H^2} \\
b &=& a - L \sinh^{-1} (R_H/L)
\end{eqnarray}
Passing through the wall of dust leads not to a region of false
vacuum, but rather to another identical region of true vacuum.

The motion of individual dust particles obeys
\begin{eqnarray}
X_1 & = & \tanh \gamma \sin \phi~X_0~, \nonumber \\
X_2 & = & \tanh \gamma \cos \phi~X_0~, \\
X_3 & = & b+L \sinh^{-1}\left(\frac{X_0}{L \cosh \gamma}\right)~, \nonumber
\end{eqnarray}
where $\gamma$ and $\phi$ label the worldline.  Their asymptotic
velocity is given by
\begin{align}
\dot{X_1} &= \tanh \gamma \sin \phi \nonumber \\
\dot{X_2} &= \tanh \gamma \cos \phi \\
\dot{X_3} &= {1 \over \sqrt{\cosh^2 \gamma + \left({X_0 \over
    L} \right)^2}} \to 0 \nonumber
\end{align}
where overdot signifies the derivative with respect to the Minkowski
time $X_0$.  

This shows that the dust particles are asymptotically moving away from
each other with a relative velocity which depends on their location.
As a result, the dust is not susceptible to gravitational collapse on
large scales.

\begin{figure}
\begin{center}
\includegraphics[scale = .37]{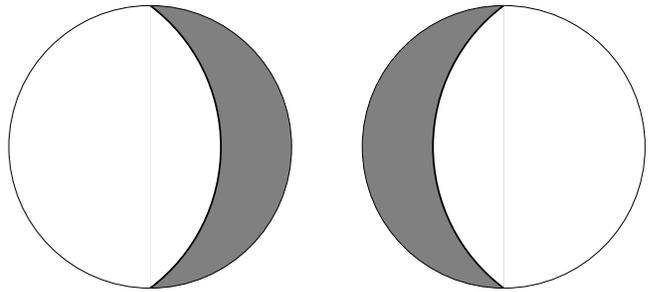}
\end{center}
\caption{A constant time slice of the two-bubble solution. The
  spatial geometry consists of two 3-hyperboloids glued across a wall
  of dust to each other. Here the 3-hyperboloids are pictured, with
  one dimension suppressed, in the Poincare disk representation. The
  two disks are to be glued along the dust wall (heavy line), and the
  shaded regions are to be thrown away.}
\label{fig-dust}
\end{figure}

\begin{figure}
\begin{center}
\includegraphics[scale = .37]{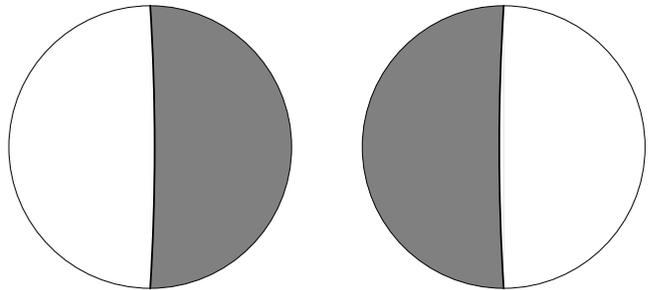}
\end{center}
\caption{At late time, the wall of dust moves to the center of the hyperboloids. In this
limit, the geometry reverts to one hyperboloid. Physically, this occurs because the dust
has diluted and no longer causes a discontinuity in the geometry across the wall. Again, 
the shaded regions are to be thrown away and the two sides glued along the heavy line.}
\label{fig-dust2}
\end{figure}

To understand the effect of the wall of dust on the open FRW universe
present in the single bubble solution, it is helpful to rewrite the
motion of the dust wall in coordinates suited to the open slicing, as
well as to the symmetries preserved by the two-bubble solution. We
write the metric in the form
\begin{equation}
ds^2 = -dt^2 + t^2 dH_3^2 \ ,
\end{equation}
with the metric on the 3-hyperboloid in the form
\begin{equation}
dH_3^2 = d\chi^2 + \cosh^2 \chi dH_2^2 \ .
\end{equation}
The $SO(1,2)$ symmetry of the solution is realized on the
2-hyperboloid. In these coordinates, at late times the dust wall obeys
\begin{equation}
\sinh \chi \approx {L \over t} \log \left( {2 t \over L} \right) \to
0 \ .
\end{equation}
On a given time slice, the spatial geometry consists of a piece of a
3-hyperboloid glued across a wall of dust to another copy of
itself. The situation is most easily pictured by suppressing one
dimension and using the Poincare disk representation of the hyperbolic
plane, as shown in figures \ref{fig-dust} and \ref{fig-dust2}.

As seen in the above equation, at late times $\chi \to 0$. The surface
$\chi=0$ is the diameter of the Poincare disk.  Figure \ref{fig-dust}
shows the dust wall at a finite value of $\chi$, while figure
\ref{fig-dust2} shows a later time, with the dust wall approaching
$\chi = 0$. Asymptotically, the two disks glue smoothly together into
a single hyperboloid, so in this sense the open FRW geometry of the
one-bubble solution is not destroyed by bubble collisions. However,
note that although on the Poincare disk the dust wall approaches the
center, the proper distance from an observer at the center ($X_1 = X_2
= X_3 = 0$) and the wall of dust grows to infinity, as is clear from
(\ref{eq-dustsoln}). It approaches the center of the disk because the
distance from the origin to the dust wall grows slower than the radius
of curvature of the hyperboloid.

We have described the domain wall collision in a frame where the two
bubbles of true vacuum are symmetric. A generic observer will
typically see one of the bubbles nucleated long before the other, and
hence much bigger than the other. Let us focus on an observer at the
center of the biggest bubble (the ``central observer''), 
keeping in mind that many observers can
be brought to the same position by using the de Sitter symmetries.
The central observer does not see the bubble collision occur in the center of
mass frame; since his bubble has had much longer to expand, its domain
wall is moving much faster than  the domain wall of the smaller bubble. 

\begin{figure}
\begin{center}
\includegraphics[scale = .37]{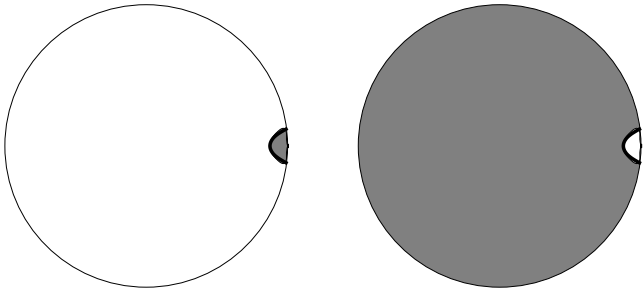}
\end{center}
\caption{A typical observer sees a boosted version of the previous
  figure, with the dust wall asymptotically moving away at a finite velocity.}
\label{fig-boosteddust}
\end{figure}

To be precise, the central observer is related to the symmetric
observer by a boost in the $X_3$ direction.  To make the bubbles very
asymmetrical requires a large boost. Now, instead of the wall of dust
coming to rest at late times, as it did in the symmetric observer's
frame, it moves away from the central observer with a finite
asymptotic velocity determined by the boost. This is shown in
figure \ref{fig-boosteddust}.

Recall that an infinite number of bubbles will eventually collide with
the ``central'' bubble. These bubbles will collide with larger and
larger boosts. The overall picture is of a central observer looking
out at the night sky, seeing walls of dust moving away with bigger and 
bigger redshift. Gravitational collapse does not occur.

\paragraph{Acknowledgements} 

We would like to thank M.~Kleban and L.~Susskind for discussions.
This work was supported by the Berkeley Center for Theoretical
Physics, by a CAREER grant of the National Science Foundation, and by
DOE grant DE-AC03-76SF00098.

\appendix

\section{Slow-roll inflation and linear perturbations}
\label{sec-slowroll}

In this appendix, we summarize slow-roll inflation and the production
of perturbations of the spatial curvature.  For a more detailed
review, see Ref.~\cite{LidLyt93}, whose notation we mostly follow.

\subsection{The flat FRW universe}
\label{sec-flat}

We work in the framework of a large, homogeneous, isotropic cosmology:
the flat FRW universe.  This is appropriate, assuming that inflation
lasts slightly longer than necessary to explain the flatness of the
observable universe.

The unperturbed metric is given by
\begin{equation}
ds^2 = -dt^2 + a(t)^2 (dx^2+dy^2+dz^2)~.
\label{eq-frw}
\end{equation}
The scale factor $a$ satisfies the Friedmann equation,
\begin{equation}
H^2 = \frac{8\pi}{3} \rho~,
\label{eq-friedmann}
\end{equation}
and the Raychaudhuri equation,
\begin{equation}
\dot{H} = -H^2 - \frac{4\pi}{3}(\rho + 3 p)~.
\label{eq-ray}
\end{equation}
The Hubble parameter $H$ is given by $\dot{a}/a$, $\rho$ is the
density, and $p$ is the pressure of matter.  

The apparent horizon is the hypersurface on which the past lightcones
centered at $\mathbf{x}=0$ have maximal area (vanishing
expansion)~\cite{RMP}.  It is located at
\begin{equation}
ax=H^{-1}~.  
\label{eq-ah}
\end{equation}
Light-rays satisfy $dt/a=\pm dx$.

In a universe dominated by matter with equation of state $p=w\rho$,
with $w\in (-1,1]$ a constant, the scale factor and density obey
\begin{equation}
a = t^{2/(3+3w)}~,~~~\rho = \rho_0 a^{-3-3w}~.
\end{equation}
A cosmological constant, $w=-1$, yields de~Sitter space (in the flat
slicing):
\begin{equation}
a = e^{Ht}~,~~~\rho = \frac{3H^2}{8\pi} = \mbox{const}~. 
\end{equation}
In this case the apparent horizon is light-like ($dt/a=-dx$) and
coincides with a cosmological event horizon.

\subsection{Inflation}
\label{sec-inf}

For simplicity, we will consider only inflationary models driven by a
single scalar field, $\phi$, with potential $V(\phi)$.  Its equation
of motion is
\begin{equation}
\ddot{\phi} + 3H \dot{\phi} = - V'(\phi)~.
\label{eq-phi}
\end{equation}
The energy density and pressure are given by
\begin{equation}
\rho = V + \frac{1}{2} \dot{\phi}^2~,~~
p = -V  + \frac{1}{2} \dot{\phi}^2~.
\label{eq-rhop}
\end{equation}
The Friedmann equations imply
\begin{equation}
\dot H = -4\pi \dot{\phi}^2
\end{equation}

We assume the slow-roll conditions $\epsilon\ll 1$ and $|\eta|\ll 1$,
where
\begin{eqnarray}
\epsilon &=& \frac{1}{16\pi} \left(\frac{V'}{V}\right)^2~, \\
\eta &=& \frac{1}{8\pi} \frac{V''}{V}~.
\end{eqnarray}
The $\ddot{\phi}$ term in Eq.~(\ref{eq-phi}) can then be neglected,
and we have
\begin{equation}
\dot{\phi} = - \frac{V'}{3H}~.
\end{equation}
That is, the motion of $\phi$ in its potential is overdamped.
Moreover, we have $\dot{\phi}^2\ll V$ in Eq.~(\ref{eq-rhop}), so
\begin{equation}
H^2 = \frac{8\pi}{3} V
\end{equation}
by the Friedmann equation (\ref{eq-friedmann}).  

Because $p\approx -\rho$ ($w\approx -1$), the evolution is similar to
de~Sitter space, 
\begin{equation}
a(t) = \exp\int H(t) dt~,\\
\end{equation}
except that the Hubble parameter slowly decreases as $\phi$ rolls
down.  The above equations imply in particular that $\dot{H}/H^2<0$,
$|\dot{H}/H^2|\ll 1$.  Hence $H$ changes only by a small fraction over
one Hubble time, $H^{-1}$, slowly increasing the size of the apparent
horizon.  From Eq.~(\ref{eq-ah}) one finds that the apparent horizon
is (barely) timelike, satisfying
\begin{equation}
\frac{dt/a}{dx} = - \left( 1+ \frac{\dot{H}}{H^2}\right)^{-1}~.
\label{eq-ahi}
\end{equation}

\subsection{Fluctuations of the inflaton}
\label{sec-flinf}

Density perturbations arise from quantum fluctuations of the inflaton
field.  Perturbing Eq.~(\ref{eq-phi}), one
obtains~\cite{Muk85,LidLyt93}
\begin{equation}
\ddot{\phi_{\mathbf k}} + 3H \dot{\phi_{\mathbf k}}
+ \left(\frac{k}{a}\right)^2 \phi_{\mathbf k} = 0~,
\label{eq-pert}
\end{equation}
where $\phi_{\mathbf k}$ is the Fourier transform of the field
perturbation $\delta\phi$:
\begin{equation}
  \delta\phi({\mathbf x},t) = (2\pi)^{-3}
\int d^3 {\mathbf k}\, \phi_{\mathbf k}(t)\, e^{i {\mathbf k} {\mathbf x}}~.
\end{equation}
Note that ${\mathbf x}$ is a comoving coordinate.  Hence, its
conjugate, ${\mathbf k}$, is a comoving momentum.  Physical scales are
given by $a {\mathbf x}$ and ${\mathbf k}/a$.

Canonical quantization yields
\begin{equation}
  \hat\phi_{\mathbf k}(t) = f_k(t) \hat a_{\mathbf k} + f^*_{-k}(t) \hat
  a^\dagger_{-\mathbf k}
  \label{eq-creann}
\end{equation}
where the creation and annihilation operators satisfy
\begin{equation} 
  [\hat a_{\mathbf k_1},\hat a^\dagger_{\mathbf k_2}] =
  (2 \pi)^3 \delta^3({\mathbf k_1} - {\mathbf k_2})~,
\end{equation}
and 
\begin{equation}
  f_k(t) = \frac{H}{(2k^3)^{1/2}} \left(i+\frac{k}{aH}\right) e^{ik/aH}
\label{eq-fk}
\end{equation}
is a solution to Eq.~(\ref{eq-pert}) which is positive frequency at
early times.  At very early times, the wavelength $a/k$ is much
smaller than the curvature radius, and one expects the field to be in
the vacuum:
\begin{equation}
  \hat a_{\mathbf k} |0\rangle=0~.
\end{equation}

As long as they are inside the horizon, the modes oscillate much like
those of a scalar field in Minkowski space.  After exiting the
horizon, the modes freeze but continue expanding, so they contain more
and more quanta.  If the number of quanta is much greater than one,
the field can be probed without altering the amplitude much.  Let us
assume, perhaps optimistically, that a stable, non-destructive
measurement becomes possible when the mode has expanded by a couple of
e-foldings after exit, when it is about a factor of $N_{\rm
  freeze}\approx 10$ larger than the horizon.

The corresponding ``freezing surface'' satisfies $ax = N_{\rm
  freeze}H^{-1}$.  Its conformal slope is $1/N_{\rm freeze}$ times the
slope of the apparent horizon, given in Eq.~(\ref{eq-ahi}):
\begin{equation}
\frac{dt/a}{dx} = - \frac{1}{N_{\rm freeze}}
\left( 1+ \frac{\dot{H}}{H^2}\right)^{-1}~.
\label{eq-freeze}
\end{equation}
Since $N_{\rm freeze}-1$ is not small, and since $\dot{H}/H^2\ll 1$,
we have $|\frac{dt/a}{dx}|< 1$.  

\begin{figure*}
\begin{center}
\includegraphics[scale = .5]{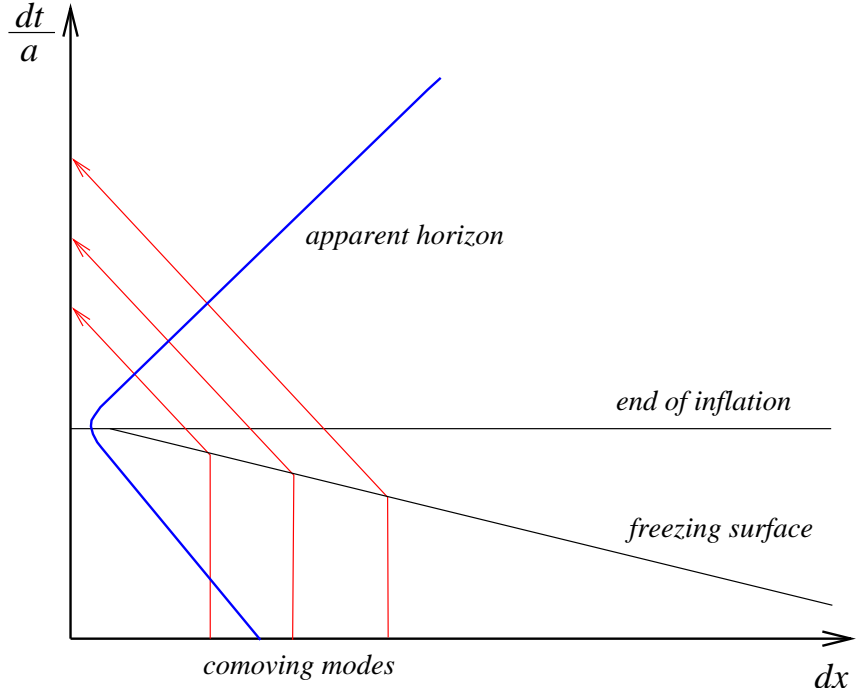}
\end{center}
\caption{The hypersurface where inflationary modes freeze out is
  spacelike.  Hence, frozen modes can be measured only after inflation
  has ended.}
\label{fig-freeze}
\end{figure*}

This shows that the freezing surface is {\em spacelike\/} and becomes
visible only after inflation ends (Fig.~\ref{fig-freeze}).

The expectation value of the Fourier amplitudes is given by
\begin{equation}
  \langle 0| \hat\phi_{\mathbf k} \hat\phi_{-{\mathbf k}}
 |0\rangle = |f_k(t)|^2~.
\end{equation}
After the mode exits the horizon ($k/a<H$), the first term in
Eq.~(\ref{eq-fk}) becomes dominant and it ceases to oscillate.  Then
\begin{equation}
  \langle 0 |\hat\phi_{\mathbf k} \hat\phi_{-{\mathbf k}} |0 \rangle 
  = \frac{H^2}{2k^3}~.
\label{eq-modes}
\end{equation}

The fluctuations are Gaussian, i.e., the phases of the Fourier
coefficients are random.  In position space, the field perturbation is
therefore given by the sum of the squared amplitudes:
\begin{equation}
\langle 0| \delta\phi(x)^2 |0\rangle = 
\frac{1}{(2\pi)^3} \int d^3k\, |f_k|^2~.
\label{eq-sqsum}
\end{equation}

In practice one does not measure a single mode, but a range of
wavelengths and orientations (say, all momenta with magnitude between
$k$ and $ek$).  The resulting perturbation is captured by the {\em
  power spectrum\/}, ${\cal P}_\phi(k)$, defined by
\begin{equation}
\langle 0| \delta\phi({\mathbf x}) \delta \phi({\mathbf {x+r}}) |0\rangle = 
\int_0^\infty {dk \over k} {\sin(kr) \over kr} {\cal P}_\phi(k) \ .
\end{equation}
The power spectrum for any other perturbation $\beta$ is defined
analogously. The power spectrum is simply related to the
mode functions $f_k$, 
\begin{equation} 
  {\cal P}_\phi(k) = \frac{k^3}{2\pi^2} |f_k|^2~.
\label{eq-pphi}
\end{equation}
After horizon exit (for $Ha/k>1$), a typical fluctuation is $H/2\pi$:
\begin{equation}
 {\cal P}_\phi = \left( \frac{H}{2\pi}\right)^2~.
\label{eq-pphi2}
\end{equation}

We are particularly interested in the behavior of the fluctuations on
the Hubble scale.  The inflaton fluctuation averaged over the Hubble
scale is given by
\begin{equation}
\delta \phi_H(t) = (2 \pi)^{-3} \int d^3 {\mathbf k} \  \sigma_t(k) \phi_k(t)
\end{equation}
where $\sigma_t(k)$ is a smearing function appropriate to averaging
over one Hubble volume at time $t$. A natural quantity to compute
which captures the buildup of the fluctuations on the Hubble scale is
\begin{equation}
\langle \left[ \delta \phi_H(t_2) - \delta \phi_H(t_1) \right]^2 \rangle~.
\end{equation}
Using the definition of $\delta \phi_H$, we find

\begin{eqnarray}
 & \langle \left[ \delta \phi_H(t_2) - \delta \phi_H(t_1) \right]^2 \rangle =  \nonumber\\
 & \langle \left[ (2 \pi)^{-3} 
 \int d^3{\mathbf k} \  \sigma_{t_2}(k) \hat\phi_{\mathbf k}(t_2) - 
 \sigma_{t_1}(k) \hat\phi_{\mathbf k}(t_1) \right]^2 \rangle
\end{eqnarray}

Using momentum conservation, we can bring the square inside the
integral. Also, we decompose the field in terms of creation and
annihilation operators as in equation (\ref{eq-creann}).  We find
\begin{eqnarray}
  &\langle \left[ \delta \phi_H(t_2) - \delta \phi_H(t_1) \right]^2 \rangle = \nonumber\\
  &(2 \pi)^{-3} \int d^3{\mathbf k} \ \left| 
    \sigma_{t_2} (k) f_k(t_2) - \sigma_{t_1}(k) f_k(t_1) \right|^2 \ .
\label{eq-modeint}
\end{eqnarray}
We consider a particularly simple smearing function whose value is 1
for modes with wavelength longer than the Hubble scale and 0
otherwise,
\begin{equation}
\sigma_t(k) =  \begin{cases} 
1  \ \  &{\rm for}  \ k<a(t) H \\
0  &{\rm for} \  k>a(t) H
\end{cases}
\end{equation}
Because the position space smearing function should integrate to 1,
the momentum space smearing function should satisfy $\sigma_t(k=0) =
1$. Our smearing function has this property. Incidentally, this
property guarantees that the quantity we are calculating is free
of the infrared divergences which appear in many computations of the
inflaton fluctuations.
Continuing with our
calculation, we note that since the smearing function only has support
on wavelengths larger than the Hubble radius, we can approximate
\begin{equation}
f_k(t) \approx i H/\sqrt{2 k^3}
\label{eq-approx}
\end{equation} 
in equation (\ref{eq-modeint}). (In
fact, this approximation becomes good for modes well outside the
horizon. We will focus on the buildup of fluctuations over many
efoldings, and we will see that the dominant contribution comes from
modes well outside the horizon.)
 With
this approximation, the integral simplifies to
\begin{eqnarray}
 & \langle \left[ \delta \phi_H(t_2) - \delta \phi_H(t_1) \right]^2 \rangle = \nonumber\\
 & {1 \over 2 \pi^2} \int_0^\infty dk \ k^2 
  \left| \sigma_{t_2}(k) - \sigma_{t_1}(k) \right|^2 {H^2 \over 2 k^3}
  \ .
\end{eqnarray}
Using the definition of the smearing function $\sigma$, the integral becomes
\begin{eqnarray}
& \langle \left[ \delta \phi_H(t_2) - \delta \phi_H(t_1) \right]^2 \rangle  \nonumber\\
& =\left({H \over 2 \pi}\right)^2  \int_{a(t_1) H}^{a(t_2) H} {dk \over k}  \nonumber\\
& =\left({H \over 2 \pi}\right)^2 \log{a(t_2) \over a(t_1)} \ .
\end{eqnarray}
The logarithmic integral above receives equal weight from every
efolding. Since we are interested in integrating over a large number
of efoldings, it is dominated by modes well outside the
horizon, so our approximation (\ref{eq-approx}) is justified.
We can rewrite our answer in a more suggestive way,
\begin{equation}
  \langle \left[ \delta \phi_H(t_2) - \delta \phi_H(t_1) \right]^2
  \rangle   =  \left({H \over 2 \pi}\right)^2 H t
\label{eq-randomwalk}
\end{equation}
This formula is correct in the limit that the number of efoldings,
$H t$, is large.
Furthermore, since the action is quadratic, the fluctuations will be
gaussian. The above formula makes clear that the variance of the
gaussian grows linearly with time, just as for a random walk.  One way
to think about it is that the inflaton field on the Hubble scale is
performing a random walk, taking a step of size $H/2 \pi$ every Hubble
time.

\subsection{Curvature perturbations}
\label{sec-curv}

Next, let us discuss how the perturbations of the inflaton field lead
to perturbations of the metric.  In semiclassical gravity, the stress
tensor entering the Einstein equation is the expectation value of the
stress tensor of quantum field theory:
\begin{equation}
R_{ab}-\frac{1}{2} R g_{ab} = 8\pi \langle T_{ab} \rangle~.
\end{equation}
In this framework the fluctuations computed above would be irrelevant,
since $\langle\delta\phi\rangle=0$.  The metric would remain
unperturbed, because the expectation value of the quantum field
remains unperturbed at linear order.

If quantum fluctuations are to shape the metric, one must go somewhat
beyond the framework of semiclassical gravity.  The point is that if
the field is measured, it takes on a typical value in the range where
its wavefunction has support.  A typical value will deviate from the
expectation value by approximately $\pm\delta\phi$.  But using a
typical value would {\em not\/} be appropriate for the ultraviolet
fluctuations of fields in flat space.  We must therefore justify this
approach carefully for the case of frozen modes of the inflaton field.

First, consider high frequency modes of $\phi$.  They are shorter than
the Hubble length, $a/k<H^{-1}$, and behave as in flat space.  By
Eq.~(\ref{eq-pphi}), the power spectrum in this regime is
\begin{equation}
{\cal P}_\phi= \frac{k^2}{2a^2} = \frac{p^2}{2}
\end{equation}
Thus a typical fluctuation is of order $p$, the physical momentum,
which is large in the ultraviolet.  These fluctuations would be seen
if such modes were measured, that is, if the field interacted with
probes sufficiently energetic to resolve small distances.  Even then,
they would not be stable, in the sense that the field would continue
to oscillate.  The interaction will have imparted significant unknown
momentum to the field, so that multiple measurements of the same range
of modes would not be strongly correlated.  Note that in the
inflationary universe, no probes are available at very high energies,
so presumably such interactions do not occur, and the quantum state of
the inflaton field remains coherent.

Modes that are larger than the Hubble scale, however,
decohere~\cite{KiePol98}.  They are probed by the
environment (at least, by gravity).  Thus, their quantum state becomes
entangled with the environment.  After tracing over unobservable
environmental degrees of freedom, coherence is lost and a density
matrix remains.  Because the amplitude remains constant while the
wavelength expands, each mode will contain a large number of quanta,
and its state will not be significantly altered by interactions.
Hence, the modes behave classically, drawn from the statistical sample
described by the density matrix.

Now let us relate the perturbations of the inflaton field to
deformations of the spatial metric.  In the unperturbed spacetime,
Eq.~(\ref{eq-frw}), the spatial metric is flat and the inflaton field
is constant ($\delta\phi=0$) on any flat slice.  With perturbations
included, the backreaction on the metric is second order, so one can
still consider flat slices, but with $\delta\phi\neq 0$.  However, in
order to evolve the effects of perturbations into the radiation and
matter dominated eras, it is more convenient to work in the comoving
gauge.  This slicing is defined by the requirement that the momentum
density of the cosmological fluid, $T_{0i}$, vanish.  During inflation
this means that $\delta\phi=0$ on comoving slices.  

Therefore, comoving slices are slightly deformed relative to the flat
slices, by a position-dependent time difference
\begin{equation}
\delta t({\mathbf x}) = - \delta\phi({\mathbf x})/\dot\phi~.
\label{eq-deltat}
\end{equation}
The three-dimensional curvature will not vanish on a comoving slice.
It is given by~\cite{LytRio98}
\begin{equation}
{\cal R}({\mathbf x}) =  H \delta t({\mathbf x})~,
\label{eq-calr}
\end{equation}
where ${\cal R}$ is ``sourced'' by the three-dimensional curvature
scalar:
\begin{equation}
-4 \nabla^2 {\cal R} = \mbox{}^3\!R~.
\end{equation}

Note that ${\cal R}$ is dimensionless.  Its physical significance is
best understood by thinking of its Fourier transform
\begin{equation} 
  {\cal R}_{\mathbf k} = \frac{1}{4} \frac{a^2}{k^2}\,
  \mbox{}^3\!R_{\mathbf k} ~.
\end{equation}
As above, one can define a power spectrum for ${\cal R} $, via
\begin{equation}
 \langle {\cal R}({\mathbf x}) {\cal R}(\mathbf{x+r}) \rangle = 
\int_0^\infty {dk \over k} {\sin(kr) \over kr} {\cal P}_{\cal R}(k) \ .
 \label{eq-lncvpb}
\end{equation}
If ${\cal P}_{\cal R}^{1/2}(k) \ll 1$, then the spatial geometry is nearly flat
on the physical distance scale $a/k$.
On the other hand,  ${\cal P}_{\cal R}^{1/2}(k)\sim 1$ indicates that
the spatial geometry is significantly curved on the physical distance
scale $a/k$.

If pressure gradients are small, different regions in a weakly
perturbed universe evolve independently.  This leads to an important
property of ${\cal P}_{\cal R}(k)$~\cite{LidLyt93}: it is
time-independent as long as the pressure gradient remains negligible
and the perturbations remain linear.  This holds during conventional
slow-roll inflation, and remains true when the mode is frozen outside
the horizon.  It also holds during matter domination since the
pressure vanishes then.  Thus ${\cal P}_{\cal R}(k)$ remains constant
for all modes that re-enter after matter has begun to dominate.  We
are interested in large scales, so we will neglect the radiation
dominated era.  Then ${\cal R}_k$ is constant for all $k$, as long as
perturbations are linear.

The curvature perturbation ${\cal R}$ is simply related to the density
contrast
\begin{equation}
\delta\equiv\frac{\delta\rho}{\rho}
\end{equation}
during the matter dominated era, through
\begin{equation}
\left(\frac{aH}{k}\right)^2\delta_{\mathbf k}  = \frac{2}{5} {\cal
  R}_{\mathbf k}~.
 \label{eq-lnmtpb}
\end{equation}
The spectral index, $n$, is defined via
\begin{equation}
n-1 = \frac{d \log {\cal P}_{\cal R}(k)}{d\log k}~.
\end{equation}
Observables such as the temperature variations in the CMB can be
computed from these quantities.  However, in the context of the
questions posed in this paper, it will serve clarity if we suppress
the indirect routes by which aspects of the geometry of the universe
are determined in practice.  Instead we will pretend, optimistically,
that we are able to measure directly the geometry of post-inflationary
comoving three-surfaces---at least, the portions contained in our past
lightcone.

We are now ready to combine results from the above discussion.  From
Eqs.~(\ref{eq-deltat}) and (\ref{eq-calr}) the spectrum of curvature
perturbations is
\begin{equation} 
  {\cal P}_{\cal R}^{1/2}(k) = \frac{H_*}{|\dot{\phi}_*|}
  {\cal P}_\phi^{1/2}(k) = \frac{H_*^2/2\pi}{V'_*/3H_*} 
  = \frac{H_*^2}{H_*'}~.
\label{eq-curvpert}
\end{equation}
The star indicates that quantities are to be evaluated at the time
when $aH/k=1$, i.e., when the mode $k$ exits the horizon during
inflation.  This is just before the quantum state decoheres and
becomes a classical perturbation of the comoving geometry.  (Because
both $H$ and $H'$ change slowly, it does not matter whether they are
evaluated a few Hubbles times too early.) Thereafter, inflation
continues and the Hubble constant varies, but as explained above, the
curvature perturbation remains constant.

\bibliographystyle{board}
\bibliography{all}
\end{document}